\documentclass[12pt,preprint]{aastex}

\renewcommand{\b}[1]{\mbox{\boldmath $#1$}}

\def\cal#1{{\cal #1}}
\def\noal#1{\noalign{\noindent\mbox{#1}}}

\def\m@th{\mathsurround=0pt}
\def\n@space{\nulldelimiterspace=0pt \m@th}
\def\biggg#1{{\mbox{$\left#1\vbox to 20.5pt{}\right.\n@space$}}}

\def\na{\nabla}

\def\p{\partial}
\def\la{\lambda}

\def\noal#1{\noalign{\noindent\mbox{#1}}}
\def\beginenum{\begin{enumerate}}
\def\endenum{\end{enumerate}}
\def\bitem{\begin{itemize}}
\def\eitem{\end{itemize}}
\def\bray{\begin{array}}
\def\eray{\end{array}}
\def\begindoc{\begin{document}}
\def\enddoc{\end{document}}
\def\bq{\begin{equation}}
\def\eq{\end{equation}}
\def\bqy{\begin{eqnarray}}
\def\eqy{\end{eqnarray}}
\def\bqyn{\begin{eqnarray*}}
\def\eqyn{\end{eqnarray*}}
\def\bc{\begin{center}}
\def\ec{\end{center}}
\def\bfll{\begin{flushleft}}
\def\efll{\end{flushleft}}
\def\bflr{\begin{flushright}}
\def\eflr{\end{flushright}}

\newcommand{\Avec}{\mbox{\boldmath $A$}}
\newcommand{\Bvec}{\mbox{\boldmath $B$}}
\newcommand{\Evec}{\mbox{\boldmath $E$}}
\newcommand{\Fvec}{\mbox{\boldmath $F$}}
\newcommand{\Gvec}{\mbox{\boldmath $G$}}
\newcommand{\Rvec}{\mbox{\boldmath $R$}}
\newcommand{\Uvec}{\mbox{\boldmath $U$}}
\newcommand{\Vvec}{\mbox{\boldmath $V$}}
\newcommand{\evec}{\mbox{\boldmath $e$}}
\newcommand{\jvec}{\mbox{\boldmath $j$}}
\newcommand{\kvec}{\mbox{\boldmath $k$}}
\newcommand{\nvec}{\mbox{\boldmath $n$}}
\newcommand{\uvec}{\mbox{\boldmath $u$}}
\newcommand{\vvec}{\mbox{\boldmath $v$}}
\newcommand{\wvec}{\mbox{\boldmath $w$}}
\newcommand{\xvec}{\mbox{\boldmath $x$}}
\newcommand{\omegavec}{\mbox{\boldmath $\omega$}}
\newcommand{\Omegavec}{\mbox{\boldmath $\Omega$}}

\begin{document}
\title{Acceleration of Plasma Flows Due to Reverse Dynamo Mechanism}
\author{Swadesh M. Mahajan\altaffilmark{1}}
\affil{Institute for Fusion Studies, The University of Texas at  Austin,
Austin, Texas 78712}
\author{Nana L. Shatashvili\altaffilmark{\ 2}
\affil{Plasma Physics Department, Tbilisi State University,
Tbilisi 380028, Georgia\\
Abdus Salam International Center for Theoretical Physics, Trieste,
Italy} Solomon V. Mikeladze and Ketevan I. Sigua }
\affil{Andronikashvili Institute of Physics, Georgian Academy of
Sciences, Georgia}
\altaffiltext{1}{\small Electronic mail: \
mahajan@mail.utexas.edu}
\altaffiltext{2}{\small Electronic mail:
\ shatash@ictp.trieste.it \hskip 0.3cm nanas@iberiapac.ge}

\clearpage

\begin{abstract}
The "reverse--dynamo" mechanism --- the amplification/generation
of fast plasma flows by micro scale (turbulent) magnetic fields
via magneto--fluid coupling is recognized and explored. It is
shown that macroscopic magnetic fields and flows are generated
simultaneously and proportionately from microscopic fields and
flows. The stronger the micro--scale driver, the stronger are the
macro--scale products. Stellar and astrophysical applications are
suggested.
\end{abstract}


\section*{}
\clearpage

The generation of macroscopic magnetic fields (primarily from
microscopic velocity fields) defines the standard "dynamo"
mechanism. The dynamo action seems to be a very pervasive
phenomenon; in fusion devices as well as in astrophysics (stellar
atmosphere, MHD jets) one sees the emergence of  macro--scale
magnetic fields from an initially turbulent system. The relaxation
observed in the Reverse Field pinches is a vivid illustration of
the dynamo in action. Search for interactions that may result in
efficient dynamo action is one of the most flourishing fields in
plasma astrophysics. The myriad phenomena taking place in the
stellar atmospheres (heating of the corona, the stellar wind etc.)
could hardly be understood without knowing the origin and nature
of the magnetic field structures weaving the corona.

The conventional dynamo theories concentrate on the generation of
macroscopic magnetic fields in charged fluids. With time the
dynamo theories have invoked more and more sophisticated physics
models --- from the kinematic to the magneto hydrodynamic (MHD)
to, more recently, the Hall MHD (HMHD) dynamo. In the latter
theories the velocity field is not specified externally (as it is
in the kinematic case) but evolves in interaction with the
magnetic field. Naturally both MHD and HMHD "dynamo" theories
encompass, in reality,  the simultaneous evolution of the magnetic
and the velocity fields. If the short--scale turbulence can
generate long--scale  magnetic fields, then under appropriate
conditions the turbulence could also generate macroscopic plasma
flows. In this context, a quotation from a recent study is rather
pertinent: the structures/magnetic elements produced by the
turbulent amplification are destroyed/dissipated even before they
are formed completely \cite{rubio,sa2005,blackman} creating
significant flows or leading to the heating.

If the process of conversion of micro--scale kinetic energy to
macro--scale magnetic energy is termed "dynamo" (D) then the
mirror image process of the conversion of micro--scale magnetic
energy to macro--scale kinetic energy could be called "reverse
dynamo" (RD). It is convenient to somewhat extend the definitions
-- the D (RD) process connotes the generation of the macroscopic
magnetic field (flow) independent of the mix of the microscopic
energy (magnetic and kinetic).

Within  the framework of a simple HMHD system, we demonstrate in
this paper  that the Dynamo  and the Reverse Dynamo processes
operate  simultaneously --- whenever a macroscopic magnetic field
is generated there is a concomitant generation of a macroscopic
plasma flow. Whether the macroscopic flow is weak
(sub--Alfv\'enic) or strong (super--Alfv\'enic) with respect to
the macroscopic field will depend on the  composition of the
turbulent energy. We shall derive the relationships between the
generated fields and the flows and discuss the conditions under
which one or the other process is dominant. In Sec.1 we display an
analytical calculation based on the conversion of micro scale
magnetic and kinetic energy into macroscopic fields and flows. In
particular, we dwell on the reverse dynamo mechanism: the
permanent dynamical feeding of the flow kinetic energy through an
interaction of the microscopic magnetic field structures with weak
flows (seed kinetic energy). In Sec.2 we illustrate that the
theoretically derived processes do indeed take place by presenting
simulation results from a general two fluid code that includes
dissipation.


\section{Theoretical Model Analysis}
\label{sec:Model}

The physical model exploited for flow generation/acceleration is
simplified HMHD -- a minimal model that entertains two interacting
scales that can be quite disparate; the macroscopic scale of the
system is generally much larger than the ion skin depth, the
intrinsic micro scale of HMHD at which ion kinetic inertia effects
become important \cite{MY-1,MMNS2,osym2,YMO}. In HMHD the ion
(${\b v}$) and electron (${\b v_e}=({\b v}-{\b j}/en)$) flow
velocities are different even  in the limit of zero electron
inertia. In its dimensionless form, HMHD comprises of
\bqy
{\p\b
b\over\p t}&=&\b\na\times\bigg[\left[\b v-\alpha_0\,\b\na\times\b
b\right]\times\b b\bigg],\\
{\p\b v\over\p t}&=&\b v\times(\b\na\times\b v)+(\b\na\times\b
b)\times\b b-\b\na\left(p+{v^{2}\over 2}\right).
\eqy
with the
standard normalizations: the density $n$ to $n_{0}$ , the magnetic
field to the some measure of the ambient field $B_0$ and
velocities to the Alfv\'{e}n velocity $V_{A0}$. We assign equal
temperatures to the electron and the protons so that the kinetic
pressure $p$ is given by: $p = p_i + p_e \simeq 2\,nT , \ T =
T_i\simeq T_e $. We note that the Hall current contributions
become significant when the dimensionless Hall coefficient
$\alpha_0 =\lambda _{i0}/R_0$ ($R_0$ -- the characteristic scale
length of a system and $\lambda _{i0}=c/\omega _{i0}$ is the
collisionless skin depth) satisfies the condition: $\alpha_0 >
\eta $, \ where $\eta $ is the inverse Lundquist number for the
plasma. For a typical solar plasma, in the corona, the
chromosphere and the transition region (TR),  this condition is
easily satisfied ($\alpha_0 $ is in the range $10^{-10} - 10^{-7}$
for densities within $(10^{14} - 10^8) \,cm^{-3}$ and $\eta =
c^2/(4\pi V_{A0}R_{\odot}\sigma ) \sim 10^{-14}$, where
$R_{\odot}$ is solar radius, $\sigma $ is the plasma
conductivity). In such circumstances, the Hall currents modifying
the dynamics of the microscopic flows and fields could have a
profound impact on the generation of macroscopic magnetic fields
\cite{mgm-1} and fast flows \cite{mnsy,msms}.

In the following analysis  \ $\alpha_0 $ \ will be absorbed by
choosing the normalizing length scale to be $\lambda_{i0}$.  Let
us now assume that our total fields are composed  of  some ambient
seed fields and fluctuations about them,
\bq
\b b=\b H+\b
b_{0}+\tilde{\b b}, \qquad\qquad \b v=\b U+{\b v_0}+\tilde{\b v}
\eq
where $\b b_{0},\ \b v_{0}$ are the equilibrium fields and $\b
H,\ \b U$ and $\tilde{\b b},\ \tilde{\b v}$ are, respectively, the
macroscopic and microscopic fluctuations.

Notice that our ambient fields are allowed to have a component at
a microscopic scale. For analytical work, we choose for the
ambient fields a special class of equilibrium solutions to Eqs.
(1-2). These  solutions, also known as the Double Beltrami (DB)
pair \cite{MY-1}, come into existence because of the interaction
of flows and fields; the Hall term is essential for their
formation. The DB configurations are known to be robust and
accessible, through a variational principle, for a variety of
conditions including inhomogeneous densities. Non constant density
cases  do display many interesting phenomena \cite{mnsy,msms}, but
the dynamo and reverse dynamo actions can be very adequately
described by the analytically tractable constant density system.
We shall, therefore, choose the following DB pair (obeying the
concomitant Bernoulli condition $\b \nabla(p_0+{\b v_0}^2/2)=const
$ \cite{osym1,osym2})
\bq
{\b b_{0}\over a}+\b\na\times\b b_{0}=\b
v_{0},\qquad \qquad \b b_{0}+\b\na\times\b v_{0}=d\b v_{0},
\eq
as
a representative ambient state. The general solution is
expressible in terms of the single Beltrami fields \ ${\b
G_{\pm}}$ that satisfies $ \b\na\times\b G(\lambda)=\lambda\b
G(\lambda)$:
\bqy
\b
b_{0}&=&C_{+}\b G_+(\la_+)+C_{-}\b G_-(\lambda_-), \\
\b v_{0}&=&(a^{-1}+\la_+)\,C_{+}\b
G_+(\la_+)+(a^{-1}+\lambda_-)\,C_{-}\b G_-(\lambda_-).
\eqy
Here
$C_{\pm}$ are the arbitrary constants and the parameters \ $a$ \
and \ $d$ \ are set by the invariants of the equilibrium system;
the magnetic helicity \ $h_{10}=\int (\b A_0\cdot \b b_0)\ d^3x$ \
and the generalized  helicity \ $h_{20}=\int (\b A_0+\b v_0)\cdot
\nabla\times({\b A_0}+ {\b v_0})d^3x$ \ \cite{MY-1,MMNS2}; here
$\b A_0$ is the vector potential of the ambient field. The inverse
scale lengths $\lambda_+$ and $\lambda_-$ are fully determined in
terms of $a$ and $d$:
$\lambda_{\pm}=\frac{1}{2}[(d-a^{-1})\pm\sqrt{( d+a^{-1})^2-4}\ ]
$. As the DB parameters $a$ and $d$ vary, $\la_{\pm}$ can range
from real to complex values of arbitrary magnitude\footnote{\ In
the analysis below we will use   \ $\lambda $ \ for the
micro--scale and \ $\mu $ \ for the macro--scale structures.}.

Our primary interest is the creation of macro fields from  the
ambient micro fields. Some what later we will assume, for
simplicity, that our zeroth order fields are wholly  at the
microscopic scale. This allows us to create a hierarchy in the
micro fields, the ambient fields are much greater than the
fluctuations at the same scale \ ($|\tilde {\b b}|\ll |\b b_{0}|,\
|\tilde{\b v}|\ll |\b v_{0}|$). Following  \cite{mgm-1}, we may
derive the following evolution equations:
\bqy
\p_{t}\b U&=&\b
U\times(\b\na\times\b
U)+\b\na\times\b H\times\b H \nonumber\\
&&\mbox{}+\left<\b v_{0}\times(\b\na\times\tilde{\b v})+\tilde{\b
v}\times(\b\na\times\b v_{0})+(\b\na\times\b b_{0})\times\tilde{\b
b}+(\b\na\times\tilde{\b b})\times\b b_{0}\right> \nonumber\\
&&\mbox{} -\left<\b\na(\b v_{0}\cdot\tilde{\b
v})\right>-\b\na\left(p+{\b U^{2}\over 2}\right),\\
 {\p\tilde{\b v}\over\p t}&=&-(\b U\cdot\b\na)\b v_{0}+(\b H\cdot\b\na)\b b_{0},\\
{\p\tilde{\b b}\over\p t}&=&(\b H\cdot\b\na)\b v_{e0}-(\b U\cdot\b\na)\b b_{0},\\
\p \b H\over \p t&=&\b\na\times\left<[\b {\tilde v}_{e}\times\b
b_{0}]+\b v_{e0}\times\tilde{\b b}\right>+\b\na\times[(\b
U-\b\na\times \b H)\times\b H],
\eqy
where the brackets $<..>$
denote the spatial averages and $\b v_{e0}=\b v_{0}-\b\na\times\b
b_{0}$. This set of equations can be regarded as a closure model
of the Hall--MHD equations, which are now general in two respects:
1) it is a closure of the full set of equations, since the
feedback of the micro--scale is consistently included in the
evolution of both ${\b H}$ and ${\b U}$;  2) the role of the Hall
current (especially in the dynamics of the micro--scale) is also
properly accounted for (see \cite{mgm-1,mgm-2} for details).

We now choose the  constants $a$ and $d$ so that the two Beltrami
scales become vastly separated (since these constants reflect the
values of the invariant helicities, it is through $a$ and $d$ that
the helicities control the final results). In the astrophysically
relevant regime of disparate scales (the size of the structure is
much greater than the ion skin depth), we shall deal with  two
extreme cases : (i) \ $a\sim d \gg 1$ , \ $(a-d)/{a\,d}\ll 1$ \
($\lambda \sim d , \ \ \mu \sim (a-d)/{a\,d}$ ), and (ii) \ $a\sim
d \ll 1$ , \ $(a-d)/{a\,d}\gg 1 \ \ (\lambda \sim (a-a^{-1}) , \ \
\mu \sim (d-a) $ ). At this time, we would like to draw the
reader's attention to the origin of scale separation in the
original equilibrium system -- it is the Hall term that imposes
the micro scale (ion skin--depth) on the macroscopic MHD
equilibrium.

Consistent with the  main objectives of this paper, we will now
assume that the original equilibrium is predominantly micro--scale
(condition applicable for many astrophysical systems), i.e, the
basic reservoir from which we will generate macro scale fields is,
indeed, at a totally different scale. Neglecting the macro scale
component altogether, the assumed equilibria becomes simpler with
the velocity and magnetic fields linearly related as
\bqy
\b v_{0}&=&\b b_{0}\left(\lambda+a^{-1}\right)\\
\noal{leading to}
\b v_{e0}&=&\b v_{0}-\b\na\times\b b_{0}=\b b_{0}\ a^{-1}\\
\dot{\tilde{\b b}}&=&\left(a^{-1}\b H-\b U\right)\cdot\b\na\b b_{0}\\
\dot{\tilde {\b v}}&=&\left(\b H-\left(\la+a^{-1}\right)\b
U\right)\cdot\b\na\b b_{0}.
\eqy
Notice the preponderance of
nonlinear terms in the evolution equations for $\b U$ and $\b H$ .
One would expect that these terms will certainly play a very
important part in the eventual saturation of the macroscopic
fields, but in the early acceleration stage when the ambient short
scale energy is much greater than the newly created macroscopic
energy, these terms will not be significant. Deferring the fully
nonlinear to a later stage, we shall limit ourselves to a
"linear'' treatment here. Neglecting the nonlinear terms and
manipulating the system of equations, we readily derive (after
"solving" for and eliminating the short scale fluctuating fields)
\bqy
\ddot{\b H } &\simeq & \left(1-{\la\over a}-{1\over a^{2}
}\right)\left<{\b\na\times(\b H\cdot\b\na)\b b_{0}\times\b
b_{0}}\right>,\\
\ddot{\b U} &\simeq &
\left<{ \left(\la+a^{-1}\right)({\b\la\dot {\tilde {\b
v}}-\b\na\times\dot {\tilde {\b
v}})-\left(\lambda+a^{-1}\right)\b\na(\b b_{0}\cdot\dot{\tilde {\b
v}})
\times\b b_{0}}}\right>\nonumber\\
&&\mbox{}-\left< (\la\dot {\tilde{\b b}}-\b\na\times\dot{\tilde{\b
b}})\times\b b_{0} \right> .
\eqy
where the spatial averages are
yet to be performed. We use the standard isotropic ABC solution of
the single Beltrami system,
\bqy
b_{0x}={b_{0}\over\sqrt
3}\left[\sin\la y+\cos\la z\right],
\nonumber\\
b_{0y}={b_{0}\over\sqrt 3}\left[\sin\la z+\cos\la x\right],
\nonumber\\
b_{0z}={b_{0}\over\sqrt 3}\left[\sin\la x+\cos\la y\right].
\eqy
to compute the spatial averages. After some tedious but
straightforward algebra, we arrive at the final acceleration
equations \bqy \ddot{\b U}&=&{\la\over 2}{b^{2}_{0}\over 3}
\b\na\times\left[\left(\left(\la+\frac{1}{a}\right)^{2}-1\right)\b
U-\la\b H\right] \eqy \bqy \ddot{\b H}&=&-\la\,{b^{2}_{0}\over
3}\left(1-{\la\over a}-{1\over a^{2}}\right)\b\na\times\b H. \eqy
where \ $b^{2}_{0}$ \ measures the ambient micro scale magnetic
energy (also the kinetic energy because of (11)). The coefficients
in these equations are determined by $a$ and $d$
($\lambda=\lambda(a,d)$).

We see that, to leading order, $\b H$ evolves independently of $\b U$ but
the reverse is not true: the evolution of $\b U$ does require knowledge
of $\b H$.

In  the dynamo context, the Hall--currents in the micro--scale are
known to  modify  the $\alpha $ coefficient  so that it survives
the standard cancellation of the kinetic and magnetic
contributions for Alfv\'enic perturbations \cite{mgm}. It is also
known that, depending on the state of the system, the  Hall effect
(by replacing the bulk kinetic helicity by the electron flow
helicity) can cause large enhancement or suppression of the dynamo
action as compared to the standard MHD \cite{mgm-2}.

Writing  (18) and (19)  as
\bq
\ddot{\b H}=-r\,(\b\na\times\b H)\
,\qquad \qquad \ddot{\b U}=\b \na \times[s\, \b U-q\b H],
\eq
where
\bq
r=\la\,{b^{2}_{0}\over 3}(1-\la \, a^{-1}-a^{-2}) \ ,
\qquad s=\la\,{b^{2}_{0}\over 6}[(\la+a^{-1})^{2}-1] \ ,  \qquad
q=\la^2\,{b^{2}_{0}\over 6} \ ,
\eq
and fourier analyzing, one
obtains
\bq
-\omega^{2}\b H=-i\,r\,(\b k\times\b H)\ , \qquad
\qquad -\omega^{2}\b U=i\b k\times(s\,\b U-q\b H).
\eq
yielding
the growth rate,
\bq
\omega^{4}=r^{2}k^{2} \ , \qquad \qquad
\omega^{2}=-|r|\,(k)\ ,
\eq
at which  $\b H$ and $\b U$ increase.
The growing  macro fields are related to one another by
\bq \b
U={q\over s+r}\ \b H.
\eq

We shall  now show  how a choice of  $a$ and $d$ fixes the relative amounts
of microscopic energy in the ambient fields and consequently  in the  nascent
macroscopic fields  $\b U$ or $\b H$. We persist with our two extreme cases:

(i) For  $\ a\sim d \gg 1$\ , the inverse micro scale $\la \sim a
\gg 1 $  implying \ $\b v_0 \sim a\,\b b_0 \gg \b b_0$, i.e, the
ambient micro--scales fields are primarily kinetic. These type of
conditions may be met in stellar photospheres, where the turbulent
velocity field at some stage can be dominant although some $\b b_0
$ \ is present as well. For these parameters, it can be easily
seen that the generated macro--fields have precisely the opposite
ordering, \ $\b U\sim a^{-1}\b H \ll \b H$. This is an example of
the straight dynamo mechanism. Micro scale fields with kinetic
dominance create, preferentially, macro scale fields that are
magnetically dominant --- super--Alfv\'enic "turbulent flows" lead
to steady flows that are equally sub--Alfv\'enic (remember we are
using Alfv\'enic units). It is extremely important, however, to
emphasize  that the dynamo effect (dominant in this regime) must
always be accompanied by the generation of macro--scale plasma
flows. This realization can have serious consequences for defining
the initial setup for the later dynamics in the stellar
atmosphere. The presence of an initial macro--scale velocity field
during the flux emergence processes is, for instance, always
guaranteed by the mechanism exposed above. The implication is that
all models of chromosphere heating / particle acceleration should
take into account the existence of macro--scale primary plasma
flows (even weak) and their self--consistent coupling (see
\cite{MMNS2,osym2,msms} and references therein).

(ii) For $\ a\sim d \ll 1\,$ the inverse micro scale \ $\la \sim
a-a^{-1} \gg 1$. Consequently  \ $\b v_0 \sim a\,\b b_0 \ll \b b_0
$, and the ambient energy is mostly magnetic. These conditions
might pertain in certain domains in the photospheres or
chromospheres, where the turbulent velocity field may exist, but
the turbulent magnetic field is the dominant component. This
micro--scale magnetically dominant initial system creates
macro--scale fields \ $\b U\sim a^{-1}\b H \gg \b H$ that are
kinetically abundant. The situation has fully reversed from the
one discussed in the previous example
--- starting from a strongly sub--Alfv\'enic turbulent flow, the
system generates a strongly super--Alfv\'enic macro--scale flow;
this mode of conversion could be called the "reverse dynamo"
mechanism. In the region of a given astrophysical system where the
fluctuating/turbulent magnetic field is initially dominant, the
magneto--fluid coupling induces efficient/significant acceleration
and part of the magnetic energy will be transferred to steady
plasma flows.  The eventual product of the "reverse dynamo"
mechanism is a steady super--Alfv\'enic flow --- a  macro flow
accompanied by a weak magnetic field (compare with \cite{bf} for a
magnetically driven dynamo. In this study magnetic field growth on
much larger scales, and significant velocity fluctuations with
finite volume averaged kinetic helicity are found). It is tempting
to stipulate that "reverse dynamo" may be the explanation for the
observations that fast flows are generally found in weak field
regions of the solar atmosphere \cite{woo2}.

This simple analysis  has led to, what we believe, are several
far-reaching results: (1) the dynamo and "reverse dynamo"
mechanisms have the same origin -- they are manifestation of the
magneto--fluid coupling; (2) The proportionality of  $\b U$ and
$\b H$ implies that they must be present simultaneously, and the
greater the macro--scale magnetic field (generated locally), the
greater the macro--scale velocity field (generated locally); (3)
the growth rate of the macro--scale fields is defined by DB
parameters (hence, by the ambient magnetic and generalized
helicities) and scales directly with the ambient turbulent energy
$\sim b_0^2\ (v_0^2)$. Thus, the larger the initial turbulent
(microscopic) magnetic energy, the stronger the acceleration of
the flow. We believe that these novel results will surely help in
advancing our understanding of the evolution of large--scale
magnetic fields and their opening up with respect to the fast
particle escape from the stellar coronae. This effect may also
have important impact on the dynamical and continuous kinetic
energy supply of plasma flows observed in various astrophysical
systems. We would add here that in this study both the initial and
final states have finite heliciies (magnetic and kinetic). The
helicity densities are dynamical parameters that evolve
self--consistently during the process of flow generation. It is
also important to notice that the end product of the reverse
dynamo action is a macroscopic flow (produced from a microscopic
helical magnetic field) while for "inverse dynamo" \cite{bf} it is
still the macroscopic magnetic field but produced from a velocity
field with helicity.

We end the analytical section by a remark on the nonlinear terms
in Eqs. (7,10) that  do not appear later. It is amazing that the
linear solution given in  Eqs. (22-24) makes the nonlinear terms
strictly zero. Thus the solution discussed in the last section is
an exact (a special class) solution of the nonlinear system and
thus remains valid even as ${\b U}$ and ${\b H}$ grow to larger
amplitudes. This interesting but peculiar property that a
basically linear solution solves the nonlinear problem pertains to
both MHD and HMHD. In MHD, for example, it manifests itself as
Walen's nonlinear Alfven wave \cite{walen1,walen2} while in HMHD
it is revealed through the recently discovered solution of
\cite{mk2}.


\section{A Simulation Example}
\label{sec:simulation}

In order to strengthen and support the conclusions of the simple
analytical model, we now present some representative results from
our 2.5 D numerical simulation of the general two-fluid equations
in Cartesian Geometry \cite{MMNS2}. For a description of the code,
the reference \cite{msms} should also be consulted. The simulation
system is somewhat different because of the existence of an
ambient embedding macroscopic field. We find that, when such a
field is present, the basic qualitatively features of the dynamo
and reverse dynamo mechanisms do not change much but the algebra
is considerably more complicated and will be presented in a longer
paper later.

The simulation system contains several effects not included in the
analysis; it has, for instance, dissipation and heat flux in
addition to the vorticity and the Hall terms.The plasma is taken
to be compressible and embedded in a gravitational field; this
provides an extra possibility for micro--scale structure creation.
Transport coefficients for heat conduction and viscosity are taken
from \cite{braginski}.

The simulation presented here deals with the trapping and
amplification of a primary flow impinging on a single closed--line
structure. The choice of initial conditions is guided by the
observational evidence \cite{A1,woo2} of the self--consistent
process of acceleration and trapping/heating of plasma particles
in the finely structured solar atmosphere. The simulation begins
with a weak  symmetric up--flow (initially Gaussian, \ ${\bf
|V|}_{0max}\ll C_{s0}$, \ where $C_{s0}$ is an initial sound
velocity) with its peak located in the central region of a single
closed magnetic field structure (location of field maximum
$B_{0z}=100\,G$ -- upper plot of Fig.1 for the vector potential
(flux function) defining the 2D arcade). This field was assumed to
be initially uniform in time. The magnetic field is represented as
: \ ${\bf B} = {\bf \nabla \times A} + B_z\,{\bf \hat z}$ \  with
\ ${\bf A}(0; A_y; 0); \ {\bf b}={\bf B}/B_{0z}; \ b_x(t,x,z\neq
0)\neq 0$. From numerous runs on the flow--field evolution, we
have chosen to display the results corresponding to the following
initial and boundary flow parameters: \ $V_{0max}(x_o,
z=0)=V_{0z}=2.18\cdot 10^5\,cm/s; \ n_{0max}=10^{12}\,cm^{-3}; \
T(x,z=0)=const=T_0 =10\,eV$. \ The background plasma density is \
$n_{bg}=0.2n_{0max}$. \ In simulations \ $n(x,z,t=0)=n/n_{0max}$ \
is an exponentially decreasing function of $z$. Experience was a
guide to for imposing the following boundary condition, \
${\partial}_x{\mathcal K}(x=\pm \infty,z,t)=0$ \ which was used
with sufficiently high accuracy for all parameters \ ${\mathcal
K}({\bf A},T,{\bf V},{\bf B},n)$ . The initial velocity field has
a pulse--like distribution (middle and lower plots of Fig.1) with
a time duration $t_0=100\,s$.

It is found that:\\
(1) the acceleration is significant in the vicinity of the
magnetic field--maximum (originally present or newly created
during the evolution) with strong deformation of field lines and
energy re--distribution due to magneto--fluid coupling and
dissipative effects.\\
{\bf (2)} Initially, a part of the flow is trapped in the maximum
field localization area, accumulated, cooled and accelerated
(plots corresponding to $t=100\,s$ in Fig.2). The accelerated flow
reaches speeds greater than $100$\,km/s  in less than $100$\,s (in
agreement with recent observations
\cite{schrijver,ami1,ryta} and references therein). \\
(3) After this stage the flow passes through a series of
quasi--equilibria. In this relatively extended era ($\sim
1000\,s$) of stochastic/oscilating  acceleration, the intermittent
flows continuously acquire energy  (see Fig.3 for the flow kinetic
and magnetic energy maxima and also Fig.2 results at
$t=1000\,s$). \\
(4) The flow starts to accelerate again (Fig.3(a-c) for the
velocity field evolution). This process is completely consistent
with the analytical prediction; the acceleration is highest in the
strong field regions  (newly generated, Fig.2). At this moment the
accelerated daughter flows (macro--scale) are decoupled from the
mother flow carrying currents and modifying the initial arcade
field creating new $b_{max}$ localization areas that span the
region between  $\lesssim 0.05\,R_s$ and $\sim 0.01\,R_s$ from the
interaction surface.

The extensive simulation runs also  show that when dissipation is
present, the hall term (proportion to\ $\alpha_0$), through the
mediation of micro--scale physics, plays a crucial role in the
acceleration/heating processes. The existence of initial fast
acceleration in the region of maximum localization of the original
magnetic field, and the creation of  new areas of macro--scale
magnetic field localization (Fig.2, panel for $A_y$) with
simultaneous transfer of the magnetic energy (oscillatory,
micro--scale) to flow kinetic energy (Fig.2, panel for $|\b V|$
and Fig.3 results) are manifestations of the combined effects of
the dynamo and reverse dynamo phenomena. The maintenance of
quasi--steady flows for rather significant period is also an
effect of the continuous energy supply from fluctuations (due to
the dissipative, Hall and vorticity effects). These flows are
likely to provide a very important  input element for
understanding the finely structured atmospheres with their
richness of dynamical structures as well as for the mechanisms of
heating, and  possible escape of plasmas.

Notice, that in the simulation the actual magnetic and generalized
helicity densities are dynamical parameters. Thus even if they are
not in the required range initially, their evolution could bring
them in the range where they could satisfy conditions needed to
efficiently generate flows. The required conditions could be met
at several stages. This could, perhaps, explain the existence of
several phases of acceleration. Dissipation effects could play a
fundamental role in setting up these distinct stages; it could,
for example, modify the generalized vorticity that will finally
lead to a modification of field lines  and even to  the creation
of micro scales (shocks or fast fluctuations).


\section{Conclusions and Acknowledgments}
\label{sec:conclusions}

From an analysis of  the two--fluid equations, we have extracted,
in this paper, the "reverse--dynamo" mechanism --- the
amplification/generation of fast plasma flows in astrophysical
systems with initial turbulent (micro scale) magnetic fields. This
process is simultaneous with, and complimentary to the highly
explored dynamo mechanism. It is found (both analytically and
numerically) that the generation of macro--scale flows is an
essential   consequence of the magneto--fluid coupling, and is
independent of  the initial and boundary conditions. The
generation of macro scale magnetic fields and flows goes hand in
hand; the greater the macro--scale magnetic field (generated
locally) the greater the macro--scale velocity field (generated
locally). The acceleration due to the reverse dynamo is directly
proportional to the initial turbulent magnetic energy. When the
microscopic magnetic field is initially dominant, a major part of
its energy transforms to macro--scale flow energy; a weak
macro--scale magnetic field is generated along with.

The reverse dynamo mechanism, providing an unfailing source for
macro--scale plasma flows, is likely to be an important mechanism
for understanding a host of phenomena in astrophysical systems.

\bigskip

Authors would like to thank Dr. R. Miklaszewski for helpful
suggestions for the improved code construction. Authors thank
Abdus Salam International Centre for Theoretical Physics, Trieste,
Italy. The study of SMM was supported by USDOE Contract
No.~DE--FG03--96ER--54366. The work of NLS, SVM, KIS was supported
by ISTC Project G-633.

\clearpage
\begin{figure}
\epsscale{.35} \plotone{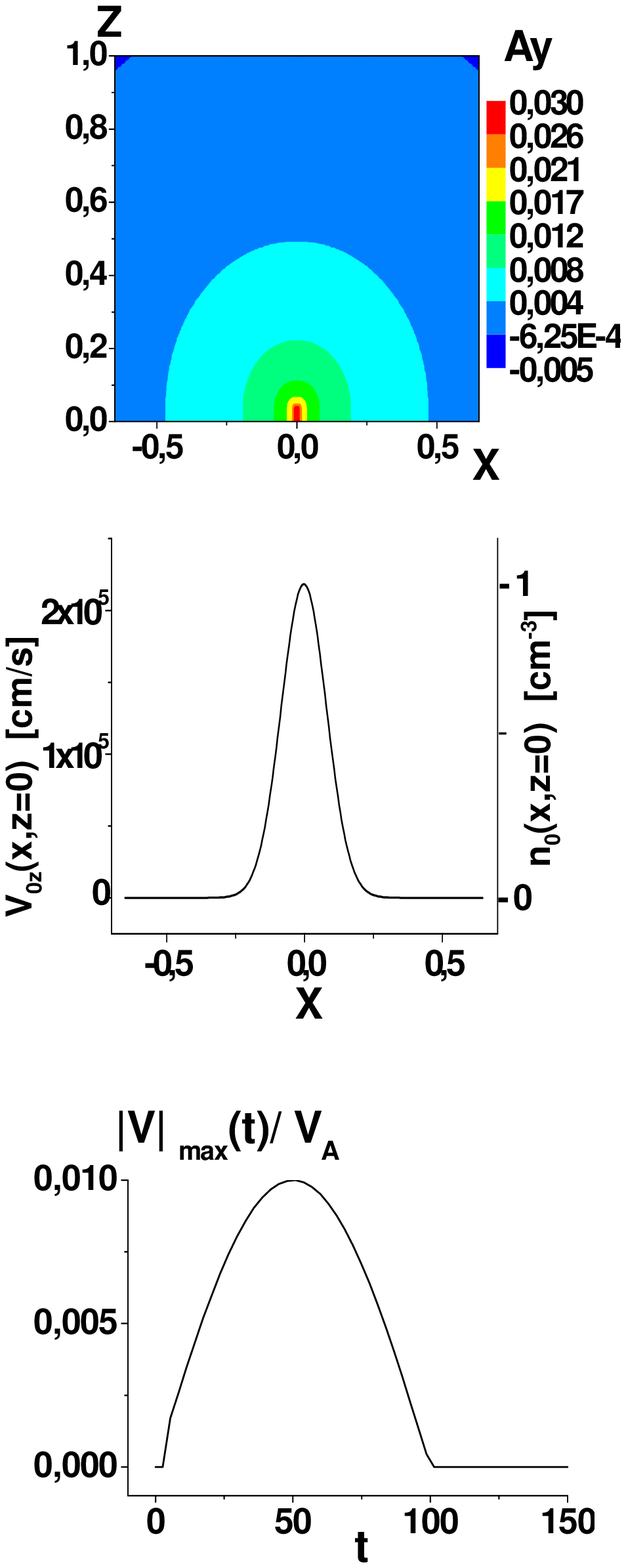} \caption{Upper plot: contour plot
for the $y$-- component of vector potential $A_y$ (flux function)
in the $x-z$ plane for an initial distribution of ambient
arcade--like magnetic field. The field has a maximum
$B_{max}(x_0=0,z_0=0)=100$\,G . Middle plot: initial symmetric
profiles of the radial velocity $V_z$, and density $n$. The
respective maxima (at $x$=0) are $\sim 2\,km/s$ and
$10^{12}\,cm^{-3}$ . Lower plot corresponds to time evolution of
initial flow: $V_z(t,z=0)=V_{0z}\,\rm{sin}(\pi t/t_0); \
V_z(t>t_0)=0; \ t_0=100\,s $ .}
\end{figure}

\clearpage
\begin{figure}
\includegraphics[angle=-90,scale=.6]{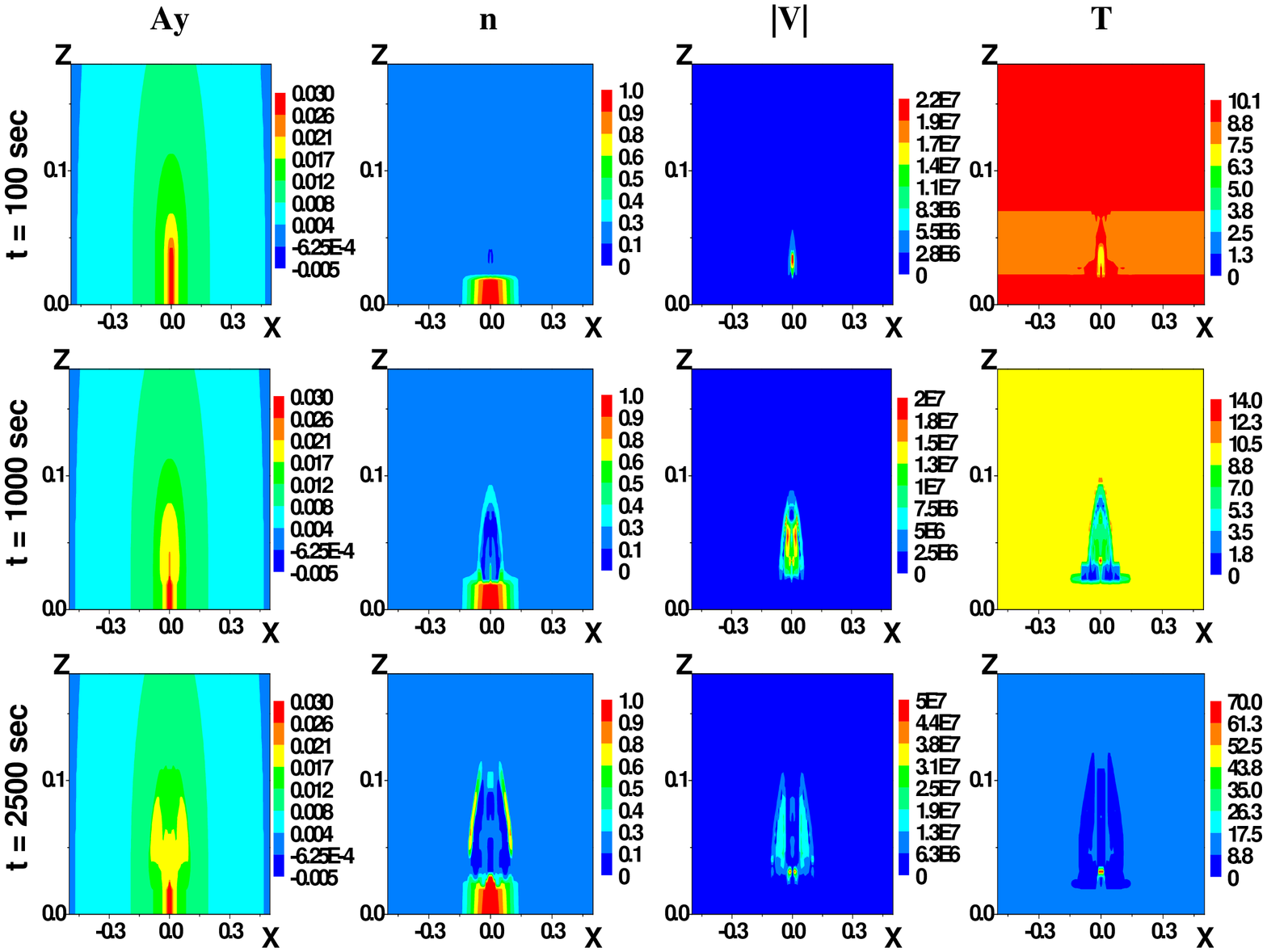}
\caption{$x-z$ contour plots at 3
time--frames: $t=100\,s; \ 1000\,s; \ 2500\,s$ \ for the dynamical
evolution of $A_y$ (first panel from the left), $n$ (second
panel), $|{\b V}|$ (third panel) and $T$ (last panel) for flow --
arcade field interaction. The realistic viscosity and heat--flux
effects as well as the Hall term ($\alpha_0=3.3\cdot 10^{-10}$)
are included in the simulation. Primary flow (type displayed in
Fig.1) is accelerated as it makes a way through the magnetic field
with an arcade--like structure (Fig.1). The primary flow, locally
sub--Alfv\'enic, is accelerated reaching significant speeds
($\gtrsim 100$\,km/s) in a very short time ($\lesssim 100$\,s).
Initially the effect is strong in the strong field region (center
of the arcade). There is a critical time ($\lesssim 1000\,s$) when
the accelerated flow bifurcates in 2; the original arcade field is
deformed correspondingly. After the bifurcation, strong magnetic
field localization areas, carrying currents, are created
symmetrically about $x=0$. Post--bifurcation daughter flows are
localized in the newly created magnetic field localization areas.
The maximum density of each daughter flow is of the order of the
density of the mother--flow. Daughter--flows have distinguishable
dimensions $\sim 0.05\,R_s$. At $t\gtrsim 1000$\,s, the velocities
reach $\sim 500$\,km/s or even greater ($\lesssim 800$\,km/s)
values. The distance from surface where it happens is \ $\gtrsim
0.01\,R_s$ . In the regions of daughter flows localization there
is a significant cooling while the nearby regions are heated.}
\end{figure}

\clearpage
\begin{figure}
\epsscale{.7} \plotone{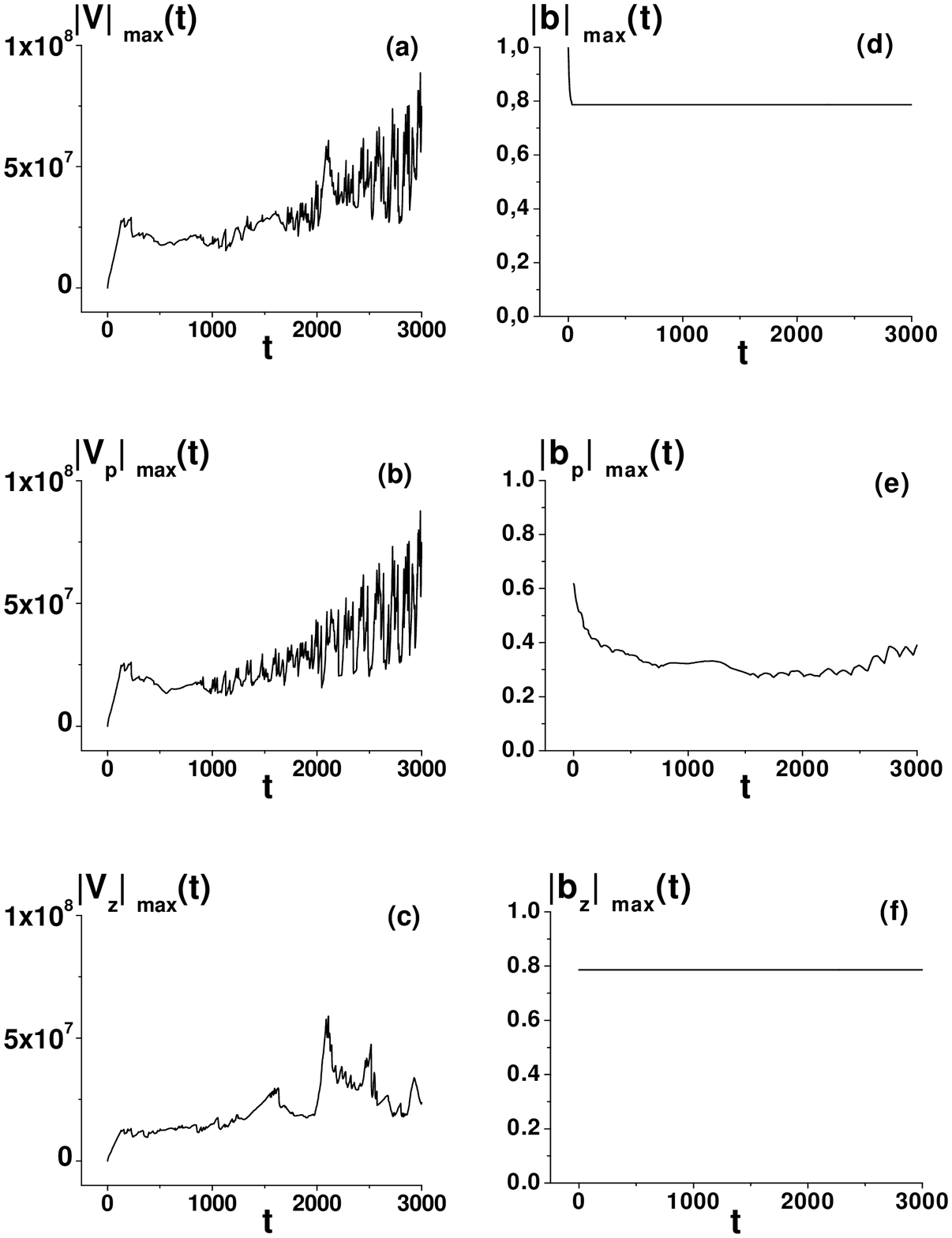} \caption{Evolution of maximum
values of  $\ |{\bf V}|, \ |\b V_p|=(V_x^2+V_y^2)^{1/2}, \ V_z$
((a--c)) and $\ |{\bf b}|, \ |\b b_p|=(b_x^2+b_y^2)^{1/2}, \ b_z$
((d--f)) in time. (a),(d) -- It is shown that much of the transfer
from magnetic field energy happens while the first and very fast
($\sim 100\,s$) acceleration stage; (e),(b) -- later, the
dissipation of perpendicular (towards height) magnetic field
fluctuations lead to the maintenance of the quasi--equilibrium
fast perpendicular flows for a period of $\sim 1000\,s$ and then
the effective acceleration of flow follows; (c),(f) -- maximum
value of magnetic field component along height is not changed and
radial component of velocity field dissipates effectively. It
should be emphasized that these maximum values of both field
parameters change the localization dynamically and follow the
relationship found analytically -- fast flows (see Fig.2) are
observed in the regions of macro scale magnetic field maximum
localization (initially given or later generated).}
\end{figure}
\end{document}